\documentclass[a4paper,fleqn,usenatbib]{mnras}

\usepackage{newtxtext,newtxmath}

\usepackage[T1]{fontenc}
\usepackage{ae,aecompl}

\usepackage{graphicx}	
\usepackage{amsmath}	
\usepackage{amssymb}	
\usepackage{ulem}	

\title[Radio transients from newborn black holes]
{Radio transients from newborn black holes}
\author[K. Kashiyama,  K. Hotokezaka, and K. Murase]
{
Kazumi Kashiyama$^{1,2}$\thanks{E-mail:kashiyama@phys.s.u-tokyo.ac.jp}, 
Kenta Hotokezaka$^{3,4}$, 
and Kohta Murase$^{5,6,7,8}$\\
$^{1}$Department of Physics, the University of Tokyo, Bunkyo, Tokyo 113-0033, Japan\\
$^{2}$Research Center for the Early Universe, the University of Tokyo, Tokyo 113-0033, Japan\\
$^{3}$Center for Computational Astrophysics, Flatiron Institute, 162 5th Ave, New York, NY 10010, USA\\
$^{4}$Department of Astrophysical Sciences, Princeton University, Princeton, NJ 08544, USA\\
$^{5}$Department of Physics, The Pennsylvania State University, University Park, PA 16802, USA\\
$^{6}$Department of Astronomy \& Astrophysics, The Pennsylvania State University, University Park, PA 16802, USA\\
$^{7}$Center for Particle and Gravitational Astrophysics, The Pennsylvania State University, University Park, PA 16802, USA\\
$^{8}$Center for Gravitational Physics, Yukawa Institute for Theoretical Physics, Kyoto University, Kyoto 606-8502, Japan
}

\date{Accepted XXX. Received YYY; in original form ZZZ}

\pubyear{2017}

\begin{document}
\label{firstpage}
\pagerange{\pageref{firstpage}--\pageref{lastpage}}
\maketitle

\begin{abstract}
We consider radio emission from a newborn black hole (BH), which is accompanied by a mini-disk with a mass of $\lesssim M_\odot$. 
Such a disk can be formed from an outer edge of the progenitor's envelope, especially for metal-poor massive stars and/or massive stars in close binaries.  
The disk accretion rate is typically super-Eddington and an ultrafast outflow with a velocity of $\sim 0.1\mbox{-}0.3\,c$ will be launched into the circumstellar medium.
The outflow forms a collisionless shock, and electrons are accelerated and emit synchrotron emission in radio bands with a flux of $\sim 10^{26-30} \ \rm erg \ s^{-1} \ Hz^{-1}$ days to decades after the BH formation. 
The model predicts not only a fast UV/optical transient but also quasi-simultaneous inverse-Compton X-ray emission $\sim$ a few days after the BH formation, and the discovery of the radio counterpart with coordinated searches will enable us to identify this type of transients. 
The occurrence rate can be $0.1-10 \ \%$ of the core-collapse supernova rate, which makes them a promising target of dedicated radio observations such as the Jansky VLA  Sky Survey.
\end{abstract}


\section{Introduction}
\begin{figure*}
\centering
\includegraphics[width=170mm]{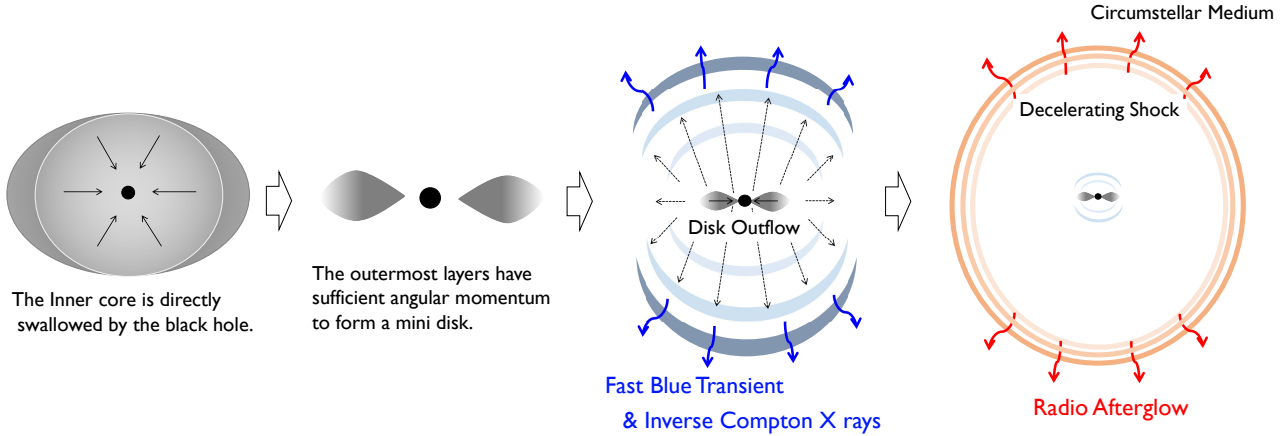}
\caption{
Schematic picture of black hole formation that is accompanied by a super-Eddington accretion disk and its electromagnetic transients.
}
\label{fig:schem}
\end{figure*}

Identification of black-hole (BH) formation events with their progenitor stars is eagerly anticipated as a missing link of massive star evolution 
and a primary target of multi-messenger time-domain astronomy. 
The recent discovery of coalescing BH binaries by the advanced LIGO/VIRGO collaboration has opened a new window of the BH astronomy~\citep{LIGO16a,LIGO16b,GW170104,GW170814}. 
Successive detections of gravitational waves will provide a fair information about the BH population in the Universe. 

Electromagnetic (EM) emission associated with BH formation should be sensitive to angular momentum profile (and magnetic field) of the progenitors~\citep[e.g.,][]{Kochanek_et_al_2008}. 
If an entire progenitor has a negligible rotation, which may be the most abundant case, almost all the stellar envelope is just swallowed by a newborn BH. 
Even in this case, a spherical mass ejection driven by neutrino mass loss in the proto-neutron star phase can produce dim electromagnetic signatures
\citep{Nadezhin_1980,Lovegrove_Woosley_2013,Piro_2013,Fernandez+18}. Some candidates have been detected~\citep{Gerke+15,Adams+17}. 
On the other hand, if a progenitor has an extremely fast rotation, a massive accretion disk is formed from the progenitor inner core just after the BH formation launching a powerful jet and disk wind, which has been considered as a central engine of the most energetic EM events, i.e., gamma-ray bursts~\citep[GRBs; e.g.,][]{MacFadyen_Woosley_1999} and (super-)luminous supernovae~\citep[SNe; e.g.,][]{Dexter_Kasen_2013}. 
Given the observed rate, such energetic events are most likely subdominant as a pathway to the BH formation. 

An intermediate case, in which only a small fraction of the progenitor outer envelope has a sufficient angular momentum to form a ``mini-disk'' (see Fig. \ref{fig:schem}), has also been of great interest. 
The BH and accretion disk system may still launch a relativistic jet, which could produce lower-power gamma-ray transients than the observed GRBs~\citep{Woosley_Heger_2012}. 
Moreover, since the accretion occurs typically with a super-Eddington rate, up to $\gtrsim 10 \ \%$ of the disk can be ejected as an outflow with a velocity of $\gtrsim 0.1 \ c$~\citep[e.g.,][]{Ohsuga_et_al_2005,Jiang_et_al_2014,Sadowski_et_al_2014}. 
Cooling emission of the ultrafast outflow will be observed as fast luminous blue transients~\citep{Kashiyama_Quataert_2015}, 
which might have already been detected~\citep[e.g.,][]{Drout_et_al_2014,Tanaka+16}. 
Related UV/optical transients with more kinetic energy have been expected at the birth of double BH binaries, which may lead to GW 150914-like merger events in $\sim1-10$~Gyr~\citep{Kimura+17}.
In general, UV/optical transients with a duration of a few days or shorter will be explored more efficiently in the coming years. 
Multi-wavelength observations will be crucial to distinguish the newborn BH scenario from alternative models. 

In this paper, we study the afterglow emission of ultrafast disk-driven outflows from newborn BHs.
At a strong collisionless shock formed at the interface of the outflow and circumstellar medium, magnetic-field amplification and particle acceleration will occur, leading to synchrotron radiation from accelerated electrons.
The radio signal may be the most promising and might have been already detected as a possible new class of radio transients, e.g., Cyg A-2~\citep{Perley_et_al_17}. 
Our scenario also predicts inverse-Compton (IC) X-ray emission in the same time window as the fast blue transient. By combining the UV/optical and X-ray signals, the radio afterglow can be used for identifying and probing this type of BH formation. 

This paper is organized as follows. 
In Sec. \ref{sec:model}, we show the setup and model for calculating the afterglow emission.
The results are shown in Sec. \ref{sec:result} and its implications are discussed in Sec. \ref{sec:discussion}. 
We use the notation $Q = 10^{x}Q_x$ in CGS units, unless otherwise noted.

\section{Setup and model}\label{sec:model}

In this section, we first describe progenitor stars assumed in our scenario, with reviewing the diversity of BH formation (Sec. \ref{sec:progenitor}). 
Then, we show properties of accretion-disk outflows (Sec. \ref{sec:outflow}) and the surrounding circumstellar medium (Sec. \ref{sec:cms}). 
The dynamics and microphysics of shocks between the outflow and circumstellar medium are modeled in Sec. \ref{sec:dynamics} and Sec. \ref{sec:micro}. 
Basic properties of the associated synchrotron emission are described in Sec. \ref{sec:sync}.
A schematic picture of our scenario is shown in Fig. \ref{fig:schem}. 

\subsection{Progenitor stars}\label{sec:progenitor}

\begin{figure}
\centering
\includegraphics[width=90mm]{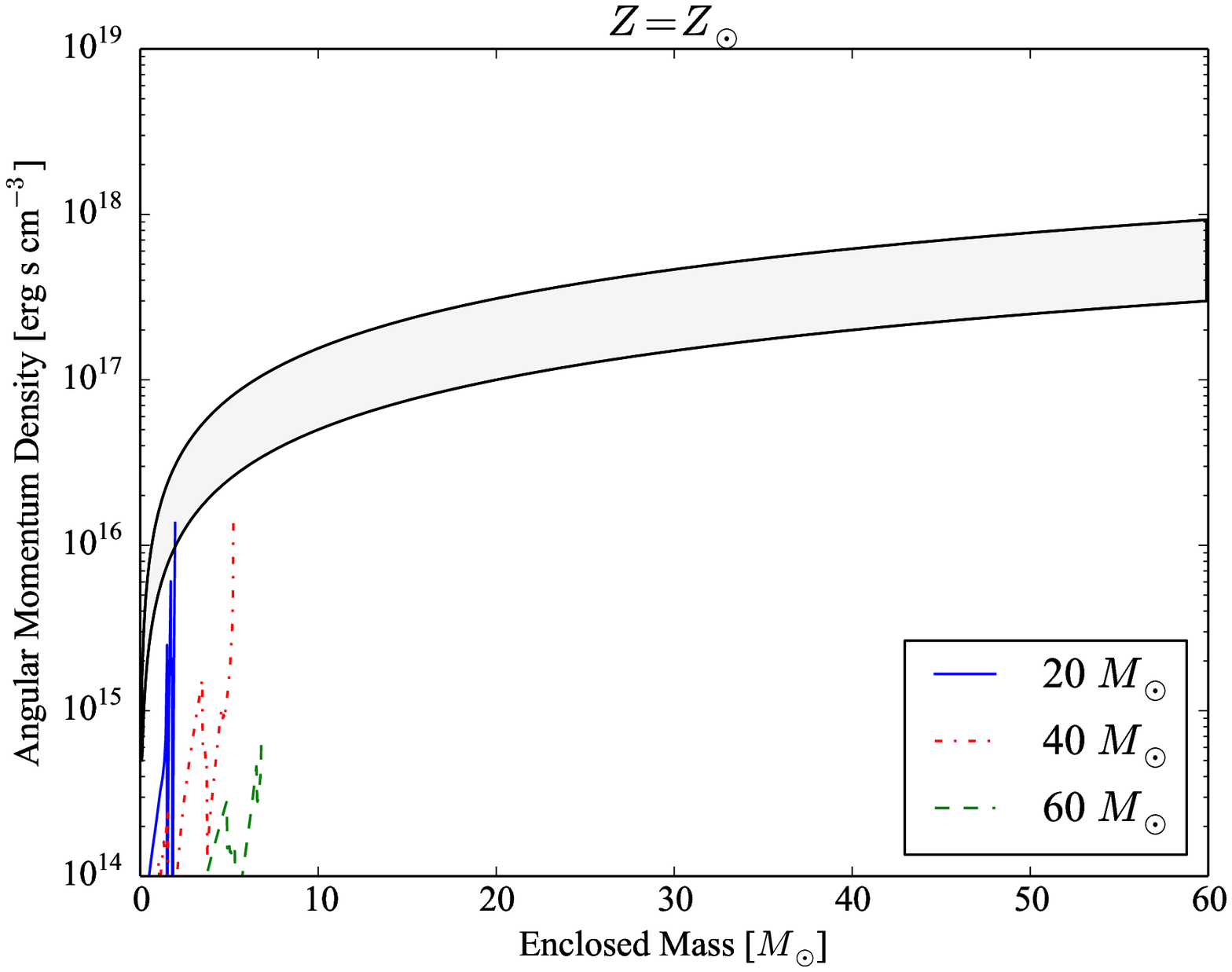}
\includegraphics[width=90mm]{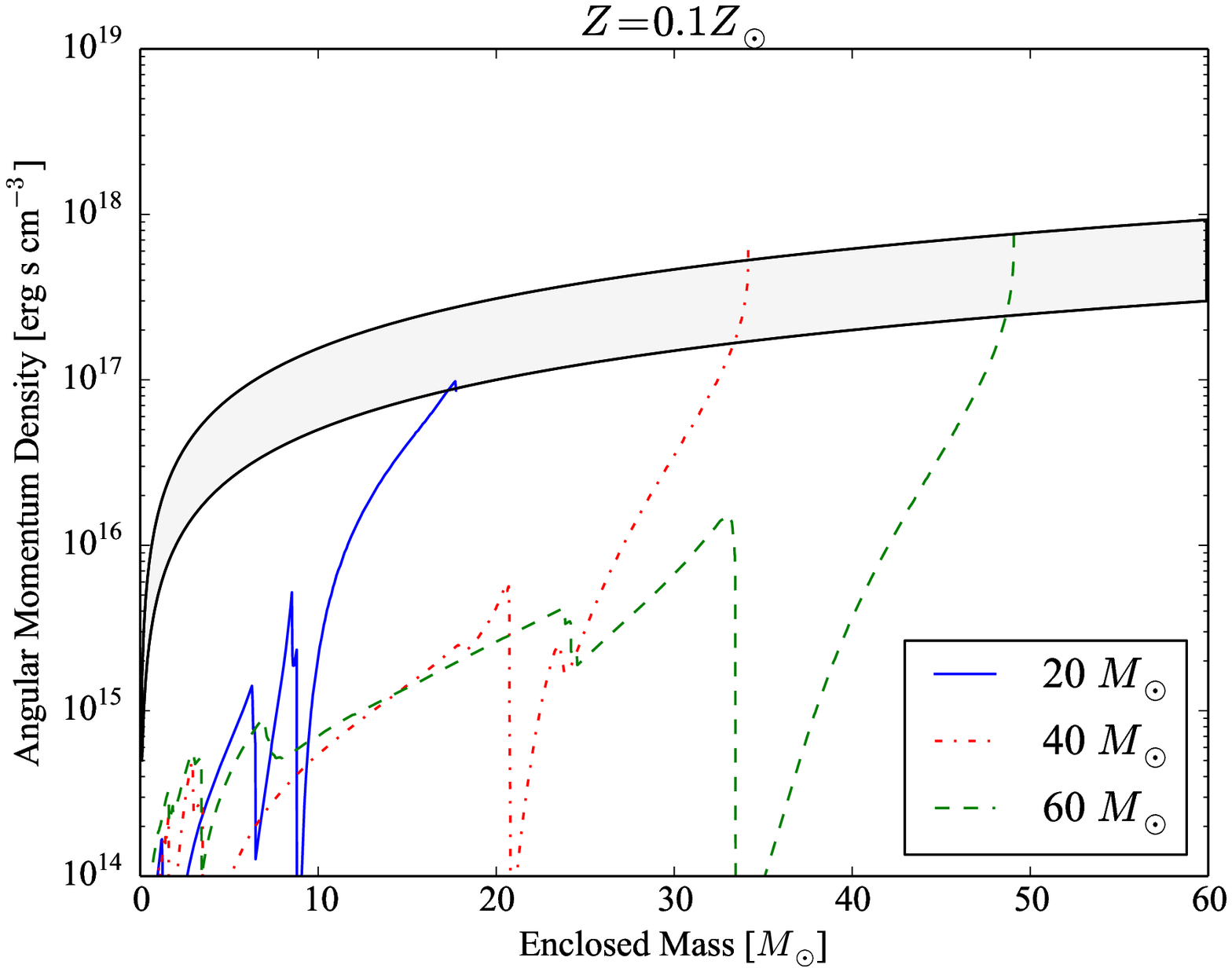}
\includegraphics[width=90mm]{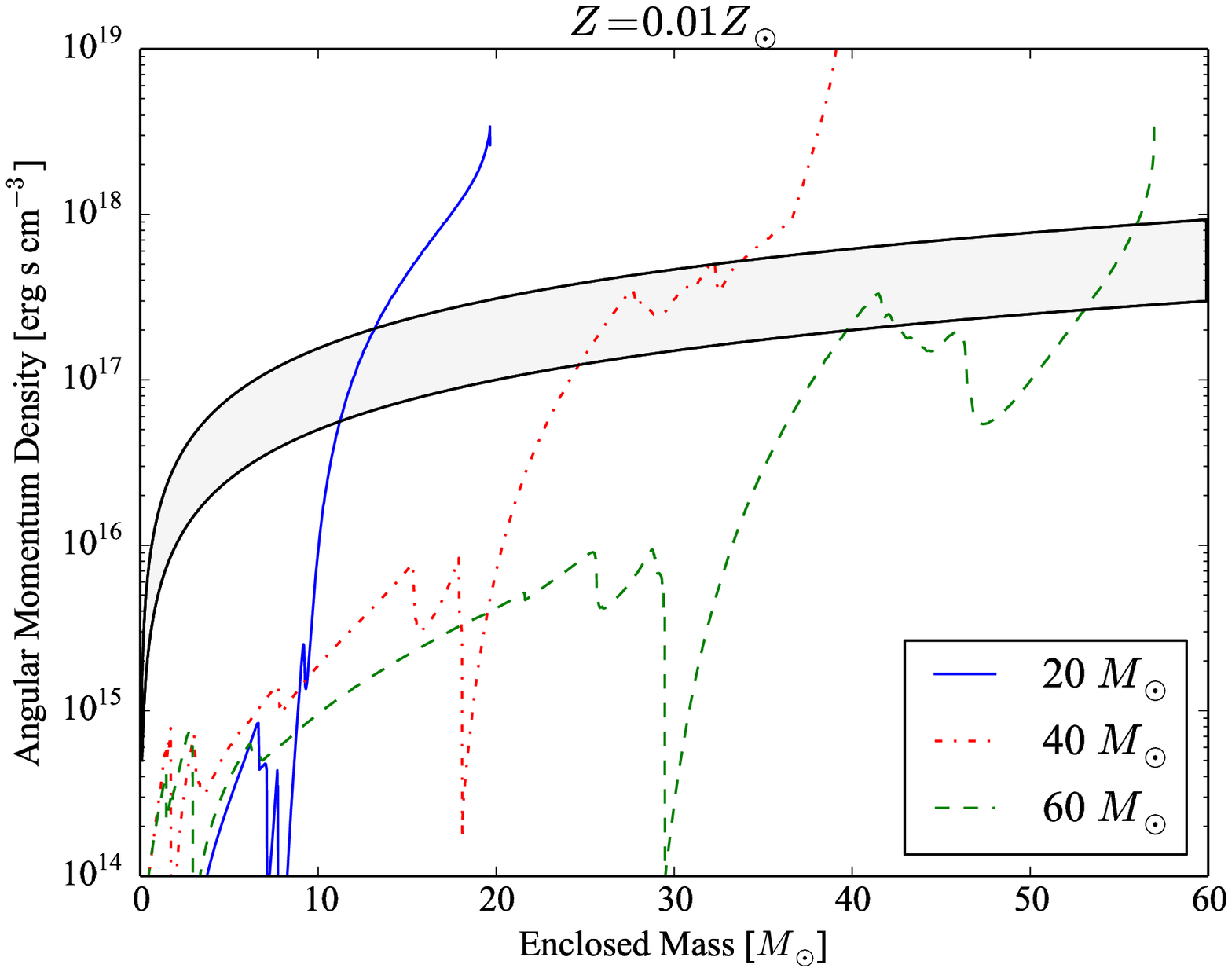}
\caption{ 
Angular momentum density profile of pre-collapse massive star calculated by using MESA. 
Each panel show the cases for single stars with $M_{\rm ZAMS} = 20 \ M_\odot$ (solid), $40 \ M_\odot$ (dotted-dash), and $60 \ M_\odot$ (dashed) and an initial surface rotation velocity of $200 \ \rm km \ s^{-1}$.  
The top, middle, and bottom panels correspond to the cases with $Z = 1, 0.1$ and $0.01 \ Z_\odot$. 
The shaded region indicate the threshold values for forming a disk or torus;  the upper and lower bounds are for Schwarzschild and extremal Kerr black hole, respectively.  
}\label{fig:progenitors}
\end{figure}

We consider collapsing massive stars which result in failed SNe and forming BHs.
The detailed conditions for SN explosions are still under debate. In general, stars with a more compact inner core more likely results in a failed explosion. 
The threshold of the compactness, which is customary defined by the enclosed mass at 1,000 km divided by 2.5 $M_\odot$, has been claimed to be $\xi_{2.5} > 0.2-0.4$ based on numerical simulations of SN explosion (\citealt{OConnor_Ott_2011,Ugliano_et_al_2012,Horiuchi_et_al_2014,Pejcha_Thompson_2015}, 
see also \citealt{Sukhbold_et_al_17}).
Once the SN shock is stalled, the shocked matter falls back onto the proto-neutron star, which finally collapses into a BH. 
Then, a rarefaction wave propagates outward and outer layers will accrete onto the BH successively. 

A disk or torus is formed around the BH if the accreting material has an angular momentum sufficiently larger than a threshold value at the inner-most-stable orbit.  
If the star rotates slow enough, the entire star is just swallowed by the BH. Such cases can be observed as vanishing massive stars or very weak SNe \citep[e.g.,][]{Kochanek_et_al_2008}, and are probably the dominant path to forming massive BHs. 
The event rate would be slightly smaller than the core-collapse SN rate, say $\lesssim 30 \ \%$~\citep{Gerke+15,Adams+17}. 
On the other hand, if the progenitor rotates so fast that even its inner core has a sufficiently large angular momentum, a massive accretion disk is formed just after the BH formation. The BH-disk system may launch a relativistic jet, which propagates though and finally punches out the progenitor star.
Such cases have been considered as the central engine of GRBs~\citep[e.g.,][]{{MacFadyen_Woosley_1999}} and probably rare; 
the true occurrence rate of GRBs is $\sim 0.1 \ \%$ of the core-collapse SN rate~\citep[e.g.,][]{Guetta_Della_07,Wanderman_Piran_10}, although the rate of low-luminosity GRBs may be higher~\citep{Liang:2006ci,Sun:2015bda}. 

In this paper, we consider an intermediate between the above two extremes \citep[or type III collapsars in][]{Woosley_Heger_2012}. 
Angular momentum density of a star basically increases with radius, and thus outer layers are more likely to form a disk when it collapses. 
The most conservative case is that only the outer most layers have sufficient angular momentum \citep[e.g.,][]{Woosley_Heger_2012,Perna_et_al_2014}. 
For the representative purpose, we calculate the angular-momentum-density profile of pre-collapse massive stars by using MESA with version 7624~\citep{Paxton_et_al_2011,Paxton_et_al_2013,Paxton_et_al_15}. 
We start from zero-age-main sequence (ZAMS), assuming a uniform rotation with a surface velocity of $200 \ \rm km \ s^{-1}$, which is consistent with the observed value of Galactic O-type stars~\citep[e.g.,][]{Fukuda_1982} and corresponds to $\sim$ a few $10 \%$ of the mass shedding limit. 
The evolution is calculated with the default parameter set for single massive stars with rotation. 
We stop our calculation when the infall velocity of a mass shell becomes larger than $10^8 \ \rm cm \ s^{-1}$, corresponding to the onset of core collapse. 

The angular-momentum-density profile for 9 cases are shown in Fig. \ref{fig:progenitors}; three different ZAMS masses, $M_{\rm ZAMS} = 20 \ M_\odot$ (solid), $40 \ M_\odot$ (dotted-dash), and $60 \ M_\odot$ (dashed) and three different metallicities, $Z = 1 \ Z_\odot$ (top), $0.1 \ Z_\odot$ (middle), and $0.01 \ Z_\odot$ (bottom).
Among them, the cases with $(M_{\rm ZAMS}, Z)= (20 \ M_\odot, \ 0.01 \ Z_\odot)$, $(40 \ M_\odot, \ 0.01 \ Z_\odot)$, $(60 \ M_\odot, \ 0.01 \ Z_\odot)$, and $(60 \ M_\odot, \ 0.1 \ Z_\odot)$ have compactness larger than the threshold value, $\xi_{2.5} > 0.2$, possibly resulting in BH formation. 
The first one is a red supergiant (RSG) and the latter three are blue supergiants (BSGs) in the pre-collapse phase. 
Note that our $Z = Z_\odot$ models are all Wolf-Rayet stars (WRs) in the pre-collapse phase, significantly losing their mass and angular momentum through the stellar wind. 
One can see that the angular momentum density generally increases significantly with respect to enclosed mass in the outer envelope.
This can be understood as follows. For a density profile of $\rho \propto r^{-n}$, the enclosed mass in the region scales as $M_{\rm enc} \propto r^{3-n}$. 
On the other hand, the angular momentum density is $j \propto v r \propto r^{2}$ for a uniform rotation. 
As a result, $j \propto M_{\rm enc}^{2/(3-n)}$, which significantly increases with $M_{\rm enc}$ for a radiative envelope with $n \approx 3$. 
The shaded regions in Fig. \ref{fig:progenitors} show the threshold values for a mass shell to form a disk around the BH. 
The upper and lower bounds correspond to Schwarzschild BHs, $j_{\rm Sch} = 2 \sqrt{3} \times GM_\bullet/c$, or 
\begin{equation}\label{eq:j_Sch}
j_{\rm Sch} \sim 4.7 \times 10^{17} \ {\rm cm^2 \ s^{-1}} M_{\bullet, 1.5}, 
\end{equation}
and extremal Kerr BHs, $j_{\rm Kerr} = (2/\sqrt{3}) \times GM_\bullet/c$, or 
\begin{equation}\label{eq:j_Kerr}
j_{\rm Kerr} \sim 1.5 \times 10^{17} \ {\rm cm^2 \ s^{-1}} M_{\bullet, 1.5}, 
\end{equation}
respectively.
\footnote{Note that the disk formation occurs in a rather complex manner, depending on the angular momentum density profile $j(r, \theta)$ of the collapsar; 
it also depends on the polar coordinate $\theta$, which cannot be calculated by 1D stellar evolution codes.  
The angular momentum density required to form a disk is larger than the threshold value we show in Fig. \ref{fig:progenitors} by a factor of a few~\citep{Zalamea_Beloborodov_09}.}
Here $M_\bullet = 10^{x} \ M_\odot \ M_{\bullet, x}$ is the mass of the BH. 
As for the three BH progenitor candidates in our sample, the outer envelope of up to a few $M_\odot$ has sufficient angular momentum to form a mini-disk.
The above results indicate a BH and mini-disk system is a quite natural outcome of massive star evolution, in particular, for $Z\lesssim0.1Z_{\odot}$ or BSG and RSG cases.

Similar situation could also occur for massive stars in tidally interacting close binaries even for WR stars~\citep{Woosley_Heger_2012}.
The tide can efficiently exchange the orbital and spin angular momentum of the massive stars and 
the spin can be synchronized with the orbital motion~\citep{Zahn_75}. 
The timescale for the synchronization of a massive star with a radiative envelope is given by 
\begin{equation}
t_{\rm syn} \sim 0.07 \ {\rm Myr} \ q^{-2} \left(\frac{1+q}{2}\right)^{-5/6} E_{2, -6} M_{\ast, 1.5}^{-1/2} R_{\ast, 12}^{-7} a_{12.5}^{17/2},
\end{equation}
where $0 < q < 1$ is the mass ratio of the binary, $E_2 \sim 10^{-7}-10^{-4}$ is a dimensionless quantity depending on the stellar structure~\citep{Zahn_75,Kushnir_et_al_17}, 
$M_\ast = 10^{x} \ M_\odot \ M_{\ast, x}$ is the stellar mass, $R_\ast$ is the stellar radius, and $a$ is the orbital separation. 
If the initial separation is smaller than a critical value 
\begin{equation}\label{eq:a_crit}
a_{\rm crit} \lesssim 4 \times 10^{12} \ {\rm cm} \ q^{4/17} \left(\frac{1+q}{2}\right)^{5/51} E_{2, -6}^{-2/17} M_{\ast, 1.5}^{1/17} R_{\ast, 12}^{14/17},
\end{equation}
the synchronization can occur within the main-sequence phase, i.e., $t_{\rm syn} \lesssim \rm Myr$. 
Once the synchronization is achieved, the spin period becomes $P_{\rm syn} \approx P_{\rm orb} = 2 \pi \sqrt{a^3/(GM_{\rm tot})}$, or
\begin{equation}
P_{\rm syn} \sim 4.4 \ {\rm days} \ a_{12.5}^{3/2} M_{\rm tot, 1.8}^{-1/2}.
\end{equation}
Assuming that the synchronous rotation is maintained until the core collapse, the angular-momentum density of the envelope is $j_{\rm syn} \approx 2 \pi R^2_\ast/P_{\rm syn}$, or 
\begin{equation}\label{eq:j_syn}
j_{\rm syn} \sim 10^{18} \ {\rm cm^2 \ s^{-1}} \ a_{12}^{-3/2} M_{\rm tot, 1.8}^{1/2} R_{\ast, 11}^2,
\end{equation}
which depends on the binary separation $a$. From Eqs. (\ref{eq:j_Sch}), (\ref{eq:j_Kerr}), (\ref{eq:a_crit}), and (\ref{eq:j_syn}), WRs in massive ($M_{\rm tot} \gtrsim {\rm a \ few} \ 10 \ M_\odot$) binaries with $a \lesssim6\times{10}^{11}~{\rm cm}\sim10R_\odot$, or massive BSGs with $R_\ast \lesssim 100 \ R_\odot$ in close binaries may result in the BH formation with a disk~\citep{Kimura+17}. 
Interestingly, these are the promising progenitors of BH-BH binaries coalescing within the cosmological timescale~\citep{Zaldarriaga2017,Hotokezaka+17A,Belczynski2017,Kimura+17}. In particular, \cite{Kimura+17} investigated electromagnetic emissions from disk-driven outflows from a newborn BH formed in such a tidally-locked binary system.

Based on the above considerations, the BH formation with a mini-disk may not be rare. Despite large uncertaintie, the event rate would be between those of vanishing massive stars and GRBs, i.e., $0.1-10 \%$ of the core-collapse SNe, which make them a promising target of ongoing and upcoming transient surveys. 
Indeed, since the mass accretion rate of a mini-disk is typically high and well above the Eddington rate (Eq. \ref{eq:M_acc}), various kinds of energetic transients can be expected.  
If there is a large-scale magnetic field connecting the BH and disk, a relativistic jet may be launched via the Blandford-Znajek process to produce a relatively dim GRB~\citep{Woosley_Heger_2012}. The more probable outcome is a strong radiation-driven wind, which has been confirmed by numerical simulations of the super-Eddington-accretion disk system~\citep[e.g.,][]{Ohsuga_et_al_2005,Jiang_et_al_2014,Sadowski_et_al_2014}. 
Hereafter, we focus on the observational signatures of such disk-driven outflows.  

We note that, even in the case of failed SNe, the outer most layers of the progenitor could be ejected by a weak shock driven by gravitational mass losses through neutrino emission in the proto-neutron star phase \citep{Nadezhin_1980}. 
In particular, in the case of RSGs, as one of our model with $(M_{\rm ZAMS}, Z)= (20 \ M_\odot, \ 0.01 \ Z_\odot)$, a part of the loosely bounded hydrogen envelope of a few $M_\odot$ can be ejected \citep{Lovegrove_Woosley_2013}.
Then, the outflow later launched from the disk, if any, interacts with the pre-ejected material, forming a shock where the energy is dissipated. 
The resultant emission can be luminous SNe \citep{Dexter_Kasen_2013}.
On the other hand, the effect of neutrino mass losses will be relatively irrelevant in the case of massive WRs and BSGs~\citep{Fernandez+18,Coughlin+18}.
Without significant quasi-spherical explosion in advance, this outflow can be almost directly seen. In this paper, we consider the latter case.

\subsection{Disk-driven outflows}\label{sec:outflow}
In this section, we show the properties of the outflows from the mini-disk and associated UV/optical transients.  

In a collapsing star, a material starts to free fall when a rarefaction wave arrives at it. 
If the material has a sufficient angular momentum, it circularizes before being swallowed by the BH.
In general, an inner mass shell will circularize earlier and more closer to the BH.
We here simply parameterize the effective circularization radius as $r_0 = f_r \times 2GM_\bullet/c^2$, or
\begin{equation}\label{eq:r_0}
r_0 \sim 1.1 \times 10^{8} \ {\rm cm} \ f_{r, 1} M_{\bullet, 1.5}.
\end{equation} 
The circularized materials will accrete when the angular momentum is transported via magnetorotational instability \citep[e.g.,][]{Proga_Begelman_2003a,Proga_Begelman_2003b}. 
Since the viscous time of the circularized material is much shorter than the free-fall time,  
the overall accretion rate is determined by the fallback rate, $\dot{M}_{\rm d} \approx M_{\rm d}/t_{\rm acc}$, or 
\begin{equation}\label{eq:M_acc}
\dot{M}_{\rm d} \sim 2.0 \times 10^{-5} \ M_\odot \ {\rm s^{-1}} \ M_{\rm d, 0.5} R_{*, 12.7}^{-3/2} M_{\bullet, 1.5}^{1/2}, 
\end{equation}
where $t_{\rm acc} \approx [R_*{}^3/(G M_\bullet)]^{1/2}$, or 
\begin{equation}\label{eq:t_acc}
t_{\rm acc} \sim 1.6 \times 10^{5} \ {\rm s} \ R_{*, 12.7}^{3/2} M_{\bullet, 1.5}^{-1/2}, 
\end{equation}
is the free fall timescale, $R_*$ is the radius of the outermost layer, and $M_{\rm d} = 10^{x} \ M_\odot \ M_{{\rm d}, x}$ is the total mass of the disk. 
We set $M_\bullet = 35 \ \rm M_\odot$, $R_* = 5 \times 10^{12} \ \rm cm$, and $M_{\rm d} = 3 \ \rm M_\odot$ as fiducial 
based on our progenitor model with $(M_{\rm ZAMS}, \ Z)= (40 \ M_\odot, \ 0.01 \ Z_\odot)$. 
The above accretion rate (Eq. \ref{eq:M_acc}) is much larger than the Eddington accretion rate, 
$\dot M_{\rm Edd} \sim 1.8 \times 10^{-14} \ M_\odot \ s^{-1} M_{\bullet, 1.5} \kappa_{-0.4}^{-1/2}$ where $\kappa$ is the opacity, 
but significantly smaller than those considered in the GRB model, $\dot M_{\rm GRB} \gtrsim 10^{-3} \ M_\odot \ s^{-1}$~\citep[e.g.,][]{Chen_Beloborodov_2007}. 
In this case, the disk will evolve into the so called slim disk~\citep{Abramowics+88,Beloborodov_98}.
Although a bulk of the disk material is advected into the BH, 
a non-negligible fraction ($f_{\rm out}\gtrsim 10 \%$) can be ejected as a radiation-driven outflow~\citep[e.g.,][]{Ohsuga_et_al_2005,Jiang_et_al_2014,Sadowski_et_al_2014}. 
The velocity of the outflow is approximately the escape velocity at the launching point, $v_{\rm out} \approx (2G M_{\bullet}/r_0)^{1/2}$, or 
\begin{equation}\label{eq:v_out}
v_{\rm out} \sim 9.5 \times 10^{9} \ {\rm cm \ s^{-1}}  \ f_{r, 1}^{-1/2}.
\end{equation}
If the launching radius is $10-100$ larger than the Schwarzschild radius, the outflow velocity ranges from $v_{\rm out} \sim 0.1-0.3 \ c$. 

In addition to the non-thermal afterglow emission considered in the following sections, 
we can expect the quasi-thermal emission diffusing out from the outflow as another electromagnetic signature of this type of BH formation~\citep{Kashiyama_Quataert_2015}. 
The emission is characterized by the peak time,
\begin{equation}\label{eq:t_p}
t_{\rm peak} \sim 3 \ {\rm days} \ f_{r, 1}^{1/4} f_{\rm out, -1}^{1/2} M_{\rm d, 0.5}^{1/2} \kappa_{-0.4}^{1/2},
\end{equation}
the peak temperature, 
\begin{equation}\label{eq:T_p}
T_{\rm peak} \sim 8000 \ {\rm K} \  f_{r, 1}^{1/8} f_{\rm out, -1}^{-1/4} M_{\rm d, 0.5}^{-1/4} R_{*, 12.7}^{1/8}  M_{\bullet, 1.5}^{1/8} \kappa_{-0.4}^{-1/2}, 
\end{equation}
and the peak bolometric luminosity, 
\begin{equation}\label{eq:Lbol_p}
L_{\rm bol, peak} \sim 6 \times 10^{42} \ {\rm erg \ s^{-1}} \ f_{r, 1}^{-1/2} R_{*, 12}^{1/2}  M_{\bullet, 1.5}^{1/2} \kappa_{-0.4}^{-1}, 
\end{equation}
which can be detected as fast luminous blue transients by UV/optical transient surveys.

\subsection{Circumstellar medium}\label{sec:cms}
Massive stars are typically formed in a dense star forming region and experience a strong mass loss during their evolution. 
In this paper, we simplify a circumstellar density profile as below;
\begin{equation}
\rho = 
\begin{cases}
\frac{\dot M_{\rm w}}{4 \pi v_{\rm w} r^2} & (r < r_{\rm w}), \\
m_{\rm p} n & (r > r_{\rm w}).
\end{cases}
\end{equation}
The inner region has a wind like density profile where $\dot M_{\rm w} = 10^x \ M_\odot \ {\rm yr^{-1}} \ \dot M_{{\rm w}, x}$ is the wind mass loss rate of the progenitor and $v_{\rm w}$ is the wind velocity. 
The density becomes constant at $r > r_{\rm w}$. 
The transition radius should be determined by the equilibrium between the ram pressure of the wind and the pressure of the surrounding medium, which yields 
\begin{equation}
r_{\rm w} = \left(\frac{5\dot M_{\rm w} v_{\rm w}}{24 \pi m_{\rm p} n \bar c_{\rm s}^2}\right)^{1/2} \sim 1.6 \ {\rm pc} \ \dot M_{\rm w, -5}^{1/2} v_{\rm w, 8}^{1/2} n_{0}^{-1/2} \bar c_{\rm s, 7}^{-1}.
\end{equation}
Here $\bar c_{\rm s}$ is the square mean of the sound and turbulent velocity in the constant-density region, assuming an adiabatic index of $5/3$. 
We note that $\bar c_{\rm s} \gtrsim 10 \ \rm km \ s^{-1}$ in a warm H-II region with $\sim 10^{4} \ \rm K$ and $\bar c_{\rm s} \gtrsim 100 \ \rm km \ s^{-1}$ in a hot H-II region with $\sim 10^{6} \ \rm K$.

\subsection{Shock dynamics}\label{sec:dynamics}
The disk outflow decelerates in the course of expanding in the circumstellar medium. 
We simplify the evolution of the shock velocity $v$ and radius $r$ as below;   
\begin{equation}
v = \left(\frac{M_{\rm out}}{M_{\rm out}+\delta M}\right)^{1/2} v_{\rm out},
\end{equation}
where $M_{\rm out} = 10^x \ M_\odot \ M_{{\rm out}, x}$ is the total mass of the outflow, 
\begin{equation}
\delta M = \int^r 4\pi r'^2 \rho dr',
\end{equation}
is the total shocked mass and 
\begin{equation}
r = \int^t v dt'.
\end{equation}
This corresponds to that the dynamics of the outflow is adiabatic, which is a good approximation in our case. 

The outflow initially coasts with an approximately constant velocity $v_{\rm out}$ and starts to decelerate once a comparable mass to the outflow mass is swept up. 
Assuming a wind-like density profile, this occurs at $r_{\rm dec, w} \approx M_{\rm out} v_{\rm w} / \dot M_{\rm w}$, or 
\begin{equation}
r_{\rm dec, w} \sim 10 \ {\rm pc} \ M_{\rm out, -1} \dot M_{\rm w, -5}^{-1} v_{\rm w, 8}.
\end{equation}
If $r_{\rm dec, w} < r_{\rm w}$, the deceleration starts in the wind medium and then the velocity evolves as $v \approx v_{\rm out} (t/t_{\rm d})^{-1/3}$. 
Here 
\begin{equation}
t_{\rm dec, w} \approx \frac{r_{\rm dec, w}}{v_{\rm out}} \sim 330 \ {\rm yr} \ M_{\rm out, -1} \beta_{\rm out, -1}^{-1} \dot M_{\rm w, -5}^{-1} v_{\rm w, 8},
\end{equation}
is the deceleration time and $\beta_{\rm out} = v_{\rm out}/c$. 
On the other hand, if $r_{\rm dec, w} > r_{\rm w}$, the deceleration of the outflow occurs in the uniform-density medium at a radius of $r_{\rm dec} \approx [3M_{\rm out}/(4\pi m_{\rm p} n)]^{1/3}$, or
\begin{equation}
r_{\rm dec} \sim 0.99 \ {\rm pc} \ M_{\rm out, -1}^{1/3} n_{0}^{-1/3}, 
\end{equation}
and a time of $t_{\rm dec} \approx r_{\rm dec}/v_{\rm out}$, or
\begin{equation}
t_{\rm dec} \sim 32 \ {\rm yr} \ M_{\rm out, -1}^{1/3} \beta_{\rm out, -1}^{-1} n_{0}^{-1/3}.
\end{equation}
Then, the velocity evolves as $\propto t^{-3/5}$ afterward.

\subsection{Shock microphysics}\label{sec:micro}
A collisionless shock is formed at the interface of the outflow and circumstellar medium, where magnetic fields are amplified.
The amplified magnetic field strength can be estimated to be
\begin{equation}
B =  (9 \pi \varepsilon_B \rho v^2)^{1/2} \sim 0.38 \ {\rm G} \ \varepsilon_{B, -2}^{1/2} \dot M_{\rm w, -5}^{1/2} v_{\rm w, 8}^{-1/2} t_{6}^{-1}.
\end{equation}
where $\varepsilon_B$ is the energy fraction carried by the magnetic field. 

In this paper, we set $\varepsilon_B = 0.01$ as fiducial following GRB radio afterglows~\citep{Meszaros_06} 
and explore a range of values, $10^{-3} \lesssim \varepsilon_B \lesssim 0.1$. 
Similar $\varepsilon_B$ values are inferred for supernova shocks~\citep[e.g.,][]{Chevalier_Fransson_16}. 
While an outflow propagates through a constant-density medium, 
\begin{equation}
B \sim 2.1 \ {\rm mG} \ \varepsilon_{B, -2}^{1/2} n_{0}^{1/2} \beta_{-1}.
\end{equation}
For $t > t_{\rm dec}$, the magnetic-field strength decreases with time as $B \propto t^{-3/5}$.  

Next we model the electron acceleration at the forward shock. 
The electrons are heated in a collisionless manner. 
The temperature can be estimated as 
\begin{equation}\label{eq:gam_eT}
\gamma_{\rm eT} = \frac{\zeta_{\rm eT}}{2} \frac{m_{\rm p}}{m_{\rm e}} \beta^2 \sim 2.2 \ \left(\frac{\zeta_{\rm eT}}{0.24}\right) \beta_{-1}^2,
\end{equation}
where $\zeta_{\rm eT}$ is the electron heating efficiency.
In this paper, we take $\zeta_{\rm eT} = 0.24$ as fiducial based on the recent particle-in-cell simulation of non- and trans-relativistic shock acceleration~\citep{Park+15}. 
In this case, the bulk of the electrons are accelerated up to trans-relativistic energies for $\beta \gtrsim 0.1$. 
A fraction of the electrons can be further accelerated by the diffusive shock acceleration mechanism. 
The injection energy is given by 
\begin{equation}\label{eq:gam_ei}
\gamma_{\rm ei} = \frac{\zeta_{\rm e}}{2} \frac{m_{\rm p}}{m_{\rm e}} \beta^2 \sim 3.7 \ \left(\frac{\zeta_{\rm e}}{0.4}\right) \beta_{-1}^2. 
\end{equation}
We take $\zeta_{\rm e} = 0.4$ as fiducial~\citep{Park+15}. Similar assumptions have been made for studies on outflow-driven radio transients in other contexts~\citep{Murase_et_al_16,Kimura+17}. 
In the Bohm limit, the shock acceleration timescale of electrons is $t_{\rm acc} = (20/3) \times c\gamma_{\rm e} m_{\rm e} c^2/ (eB v^2)$.
On the other hand, the synchrotron cooling timescale is $t_{\rm syn} \approx 6 \pi m_{\rm e}c/(\sigma_{\rm T} B^2 \gamma_{\rm e})$. 
The possible maximum Lorentz factor of non-thermal electrons can be given by $t_{\rm acc} = t_{\rm syn}$, which yields $\gamma_{\rm eM} = [9\pi e \beta^2/(10 \sigma_{\rm T} B)]^{1/2}$, or 
\begin{equation}\label{eq:gamma_eM_syn}
\gamma_{\rm eM} \sim 
\begin{cases}
7.3 \times 10^6 \ \varepsilon_{B, -2}^{-1/4} \dot M_{\rm w, -5}^{-1/4} v_{\rm w, 8}^{1/4}  \beta_{-1} t_{6}^{1/2} & (r < r_{\rm w}), \\
9.9 \times 10^7 \ \varepsilon_{B, -2}^{-1/4} n_{0}^{-1/4} \beta_{-1}^{-1/2} & (r > r_{\rm w}).
\end{cases}
\end{equation}
The injection spectrum of electron consists of a thermal component and a power-law component; 
\begin{equation}\label{eq:Q_e}
Q_{\rm e} = {\cal C}_{\rm th} \gamma_{\rm e} \sqrt{\gamma_{\rm e}^2 -1} \exp\left(-\frac{\gamma_{\rm e}}{\gamma_{\rm eT}}\right) + {\cal C}_{\rm nth} \gamma_{\rm e}^{-p}. 
\end{equation} 
The power-law component ranges from $\gamma_{\rm ei} \leq \gamma_{\rm e} \leq \gamma_{\rm eM}$.
The normalization factors ${\cal C}_{\rm th}$ and ${\cal C}_{\rm nth}$ are determined by fixing $f_{\rm e}$, the number fraction of electrons being shock-accelerated. In particular, we have
\begin{equation}\label{eq:e_cons}
f_e (4\pi r^2) \rho v =\int d\gamma_{\rm e} \, C_{\rm nth} \gamma_{\rm e}^{-p}.
\end{equation}
Note that for given $f_e$ and $\zeta_e$ the energy fraction carried by non-thermal electrons, $\epsilon_e$, is correspondingly determined.
After the injection, the Lorentz factor of the electrons evolves as 
\begin{equation}\label{eq:elec_spec}
\frac{\partial n_{\gamma_{\rm e}}}{\partial t} = - \frac{\partial}{\partial \gamma_{\rm e}} \left[ \left\{ \left(\frac{d\gamma_{\rm e}}{dt}\right)_{\rm syn} + \left(\frac{d\gamma_{\rm e}}{dt}\right)_{\rm ad} \right\} n_{\gamma_{\rm e}} \right] + Q_{\rm e}.
\end{equation}
The first term in the right hand side represents the adiabatic cooling. 
The synchrotron cooling is relevant for electrons with $t_{\rm syn} \leq t$, which corresponds to $\gamma_{\rm e} \geq \gamma_{\rm ec} = 6 \pi m_{\rm e} c/(\sigma_{\rm T} B^2 t)$.
The cooling Lorentz factor can be estimated as 
\begin{equation}\label{eq:gamma_ec_syn}
\gamma_{\rm ec} \sim 
\begin{cases}
5.4 \times 10^3 \ \varepsilon_{B, -2}^{-1} \dot M_{\rm w, -5}^{-1} v_{\rm w, 8} t_{6} & (r < r_{\rm w}), \\
1.8 \times 10^6 \ \varepsilon_{B, -2}^{-1} n_{0}^{-1} \beta_{-1}^{-2} t_{8}^{-1} & (r > r_{\rm w}).
\end{cases}
\end{equation}
One can see that the electrons are in slow cooling regime, i.e., $\gamma_{\rm ei} < \gamma_{\rm ec}$ for $t \gtrsim 1000 \ \rm s$.
Note that we here neglect IC cooling, which can be relevant in the early time, $t \lesssim t_{\rm peak} \sim \rm a \ few \ days$ (Eq. \ref{eq:t_p}).
See Sec. \ref{sec:discussion} for the associated IC emission.  
We should also note that, when the shock decelerates significantly, both the temperature and injection energy of the non-thermal electrons become non-relativistic and our treatment (Eqs \ref{eq:gam_eT}, \ref{eq:gam_ei}, and \ref{eq:Q_e}) is no longer valid (see e.g., \cite{Sironi_Giannios_13}). 

\subsection{Radio synchrotron emission}\label{sec:sync}
Once an electron spectrum is obtained, the synchrotron emission can be calculated. 
In this subsection, we discuss the basic properties. 
Detailed results are shown in the next section. 

A synchrotron spectrum can be characterized by the typical frequency $\nu_{\rm i} \approx \gamma_{\rm ei}^2 eB/(2 \pi m_{\rm e} c)$, or 
\begin{equation}
\nu_{\rm i} \sim 
\begin{cases}
9.2 \times 10^7 \ {\rm Hz} \ \zeta_{\rm e}^2 \varepsilon_{B, -2}^{1/2} \dot M_{\rm w, -5}^{1/2} v_{\rm w, 8}^{-1/2} \beta_{-1}^4 t_{6}^{-1} & (r < r_{\rm w}), \\ 
5.1 \times 10^5 \ {\rm Hz} \ \zeta_{\rm e}^2 \varepsilon_{B, -2}^{1/2} n_{0}^{1/2} \beta_{-1}^5 \ & (r > r_{\rm w}), 
\end{cases}
\end{equation}
the maximum frequency $\nu_{\rm i} \approx \gamma_{\rm eM}^2 eB/(2 \pi m_{\rm e} c)$, or 
\begin{equation}
\nu_{\rm M} \sim 5.8 \times 10^{19} \ {\rm Hz} \ \beta_{-1}^{2}, 
\end{equation}
and the cooling frequency $\nu_{\rm i} \approx \gamma_{\rm ec}^2 eB/(2 \pi m_{\rm e} c)$, or
\begin{equation}
\nu_{\rm c} \sim 
\begin{cases}
3.1 \times 10^{13} \ {\rm Hz} \ \varepsilon_{B, -2}^{-3/2} \dot M_{\rm w, -5}^{-3/2} v_{\rm w, 8}^{3/2} t_{6} & (r < r_{\rm w}), \\
1.9 \times 10^{16} \ {\rm Hz} \ \varepsilon_{B, -2}^{-3/2} n_{0}^{-3/2} \beta_{-1}^{-3} t_{8}^{-2} & (r > r_{\rm w}).
\end{cases}
\end{equation}
If the emitter is optically thin, the maximum synchrotron flux is $F_{\nu, \rm max} \approx [0.6 f_{\rm e}/(4 \pi d^2)] \times N_{\rm e} \times [\sqrt{3} \pi e^3 B/(m_{\rm e} c^2)]$, or
\begin{equation}
F_{\nu, \rm max} \sim 
\begin{cases}
4.9 \ {\rm mJy} \ f_{\rm e} \varepsilon_{B, -2}^{1/2} \dot M_{\rm w, -5}^{3/2} v_{\rm w, 8}^{-3/2} \beta_{-1} d_{\rm 27}^{-2} & (r < r_{\rm w}),\\ 
0.26  \ {\rm mJy} \ f_{\rm e} \varepsilon_{B, -2}^{1/2} n_{0}^{3/2} \beta_{-1}^{4} t_{8}^{3} d_{\rm 27}^{-2} & (r > r_{\rm w}), 
\end{cases}
\end{equation}
at $\nu \approx \nu_{\rm i}$.
Here $d$ is the distance to the source.
In fact, synchrotron self-absorption (SSA) is relevant in the radio band. 
The optical depth for SSA can be calculated as \citep[e.g.,][]{Ghisellini_Svensson_91} 
\begin{equation}
\tau_{\rm ssa} = r \int d\gamma_{\rm e} \frac{dn_{\rm \gamma_{\rm e}}}{d\gamma_{\rm e}} \sigma_{\rm ssa}(\nu, \gamma_{\rm e}),
\end{equation}
where 
\begin{equation}
\sigma_{\rm ssa} = \frac{1}{2m_{\rm e}\nu^2 \gamma_{\rm e} p_{\rm e}}\frac{\partial}{\partial \gamma_{\rm e}} [\gamma_{\rm e} p_{\rm e} j_{\rm syn}(\nu,\gamma_{\rm e})]
\end{equation}
is the SSA cross section with $p_{\rm e}$ being the momentum of electron and $j_{\rm syn} (\nu, \gamma_{\rm e})$ is the synchrotron emissivity. 
The synchrotron spectrum including the SSA frequency depends on the electron spectrum, which we calculate consistently with the synchrotron cooling by numerically solving Eq. (\ref{eq:elec_spec}).
\footnote{We neglect the heating of electrons by SSA.}
In our case, a relation $\nu_{\rm i} < \nu_{\rm ssa} < \nu_{\rm c} < \nu_{\rm M}$ is typically satisfied
and the radio spectrum peaks around $\nu = \nu_{\rm ssa}$.
Analytic formulae of the spectrum for single-power-law electrons are shown in Appendix. 


\section{Results}\label{sec:result}

\begin{table}
\centering
\caption{Model parameters used in Fig. \ref{fig:fbt}.}
\label{table:para}
\begin{tabular}{l*{1}{c}}
\hline
Disk outflow & \\
\hline
Total mass ($M_{\rm out}$) & $0.1 \ M_\odot$  \\
Mean velocity ($v_{\rm out}$) & $0.3 \ c$  \\
\hline
Microphysics of shock & \\
\hline
Magnetic field amplification efficiency ($\varepsilon_{\rm B}$) & $0.01$ \\
Electron heating efficiency ($\zeta_{\rm eT}$) & 0.24  \\
Electron acceleration efficiency ($\zeta_{\rm e}$) & 0.4  \\
Number fraction of electron accelerated ($f_{\rm e}$) & 0.05  \\
Power law index of non-thermal electron ($p$) & $2.0$  \\
\hline
Circumstellar medium & \\
\hline
Wind mass loss rate of progenitor ($\dot M_{\rm w}$) & $10^{-5} \ \rm M_\odot \ yr^{-1}$  \\
Wind velocity ($v_{\rm w}$) & $10^{8} \ \rm cm \ s^{-1}$  \\
Number density in uniform medium ($n$) & $1 \ \rm cm^{-3}$  \\
Effective sound velocity in uniform medium ($\bar c_{\rm s}$) & $10^{7} \ \rm cm \ s^{-1}$  \\
\hline
\end{tabular}
\end{table}

\begin{figure}
\centering
\includegraphics[width=85mm]{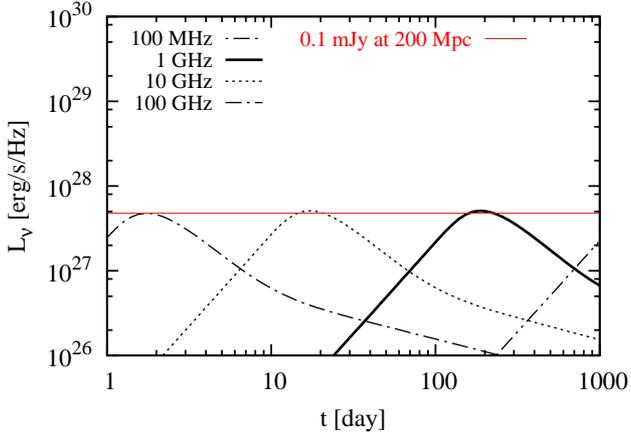}
\caption{Radio transients from a massive black hole formation with a mini-disk. The model parameters are set as in Table 1. The black lines show the light curves of different bands and the red line indicates a flux level corresponding to an observed flux of $0.1 \ \rm mJy$ from a source at $200 \ \rm Mpc$. 
}\label{fig:fbt}
\end{figure}

Fig. \ref{fig:fbt} shows light curves (from 100 MHz to 100 GHz) of a radio transient from a  BH formation with a mini-disk. 
We set the model parameters as in Table \ref{table:para}, which are consistent with the ones used for explaining the observed fast blue transients 
\citep[see Eqs. \ref{eq:t_p}-\ref{eq:Lbol_p} and][]{Kashiyama_Quataert_2015}. 
In this figure, the entire emission is originated from a shock going through a wind region.
The raising part and peak flux of the light curves are determined by SSA.
The light curves rise up as $F_{\nu} \propto t^2$ and a higher frequency band takes its peak faster, $t_{\rm peak} \propto \nu^{-1}$. 
The peak flux is independent of the frequency, $F_{\nu, \rm peak} \propto \nu^{0}$.  
The thin solid red line corresponds to $0.1 \ \rm mJy$ from a source at $200 \ \rm Mpc$. 
For example, the detection threshold of e.g., the Jansky VLA~\citep{Perley_et_al_11} and ASKAP~\citep{Johnston_et_al_08} with an integration time of $\sim$ an hour is $\sim 0.03-0.1 \ \rm mJy$. 
Thus, the radio transient can be detectable up to $\gtrsim 200 \ \rm Mpc$, which is the typical distance to fast blue transients observed so far~\citep{Drout_et_al_2014}.
We strongly encourage radio follow-up observations $\sim$ 10 to 100 days after the optical detection. 

Note that a bright radio transient with $L_\nu\sim{10}^{27}-{10}^{29}~{\rm erg}~{\rm s}^{-1}~{\rm Hz}^{-1}$ has been predicted for Type IIn SNe~\citep[see Eqs.~56 and 58 in][]{Murase_Thompson_Ofek_14}. While the radio emission will be strongly suppressed just after the photon breakout, it is detectable at late times, which is consistent with the observations of some luminous Type IIn SNe~\citep{Chandra:2015jsa}. However, the model predicts the association with long-lasting optical transients, unlike fast UV/optical transients.  Also, their contribution from secondary emission can be used for discriminating the predicted signals. 

\begin{figure}
\centering
\includegraphics[width=85mm]{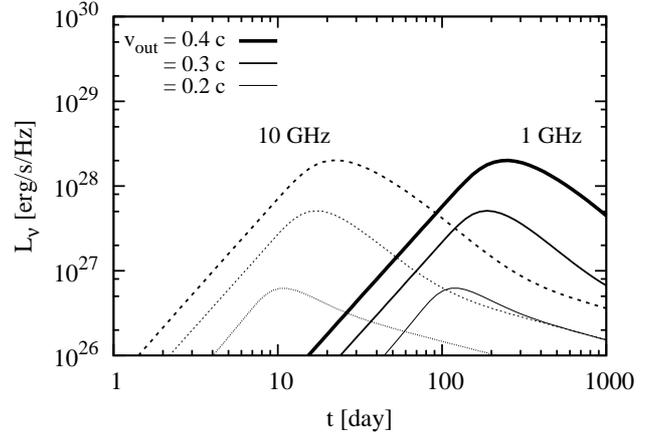}
\caption{ 
1 and 10 GHz light curves of massive black hole formation with a mini-disk. The model parameters are the same as Fig. \ref{fig:fbt} except for the velocity of disk outflow. 
}\label{fig:vout}
\end{figure}

\begin{figure}
\centering
\includegraphics[width=85mm]{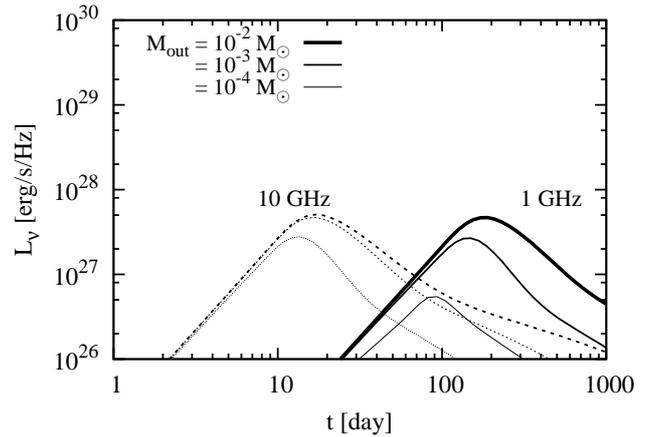}
\caption{ 
1 and 10 GHz light curve of a massive black hole formation with a mini-disk. The model parameters are the same as Fig. \ref{fig:fbt} except for the total mass of disk outflow. 
}\label{fig:Mout}
\end{figure}

In general, there should be a diversity in the parameters of the outflow from newborn BHs depending on the progenitor structure. 
Fig. \ref{fig:vout} shows the dependence of the radio emission on the outflow velocity.
We fix the other parameters as in Fig. \ref{fig:fbt}. 
A faster outflow, which can be launched from an inner radius of the disk, can give a significantly larger peak flux.
This is simply because that a faster outflow results in a larger energy dissipation rate at the shock. 
On the other hand, the peak time does not change much depending on $v_{\rm out}$, which is mainly due to that the SSA frequency is relatively insensitive to $v_{\rm out}$.
Fig. \ref{fig:Mout} shows the dependence of the radio emission on the outflow mass. 
Again we fix other parameters as in Fig. \ref{fig:fbt}. 
A larger outflow mass corresponds to a larger disk mass. 
For a sufficiently large outflow mass, the peak time and flux are determined by SSA as in Fig. \ref{fig:fbt}. 
On the other hand, for a smaller outflow mass, the deceleration of the ejecta sets in before the SSA peak especially for the lower frequency bands.
In this case, the peak of the light curve comes earlier and the peak flux becomes smaller.
By detecting such features, one can constrain the outflow mass from newborn BHs.   
To this end, a multi-band radio observation is crucial. 

\begin{figure}
\centering
\includegraphics[width=85mm]{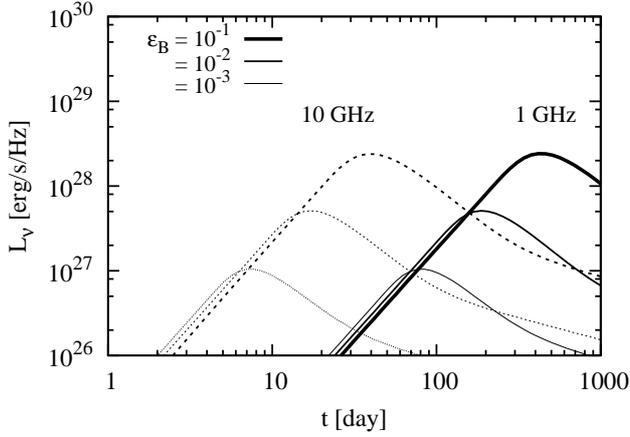}
\caption{ 
1 and 10 GHz light curve of a massive black hole formation with a mini-disk. 
The model parameters are the same as Fig. \ref{fig:fbt} except for the magnetic field energy fraction at the forward shock. 
}\label{fig:B_amp}
\end{figure}

\begin{figure}
\centering
\includegraphics[width=85mm]{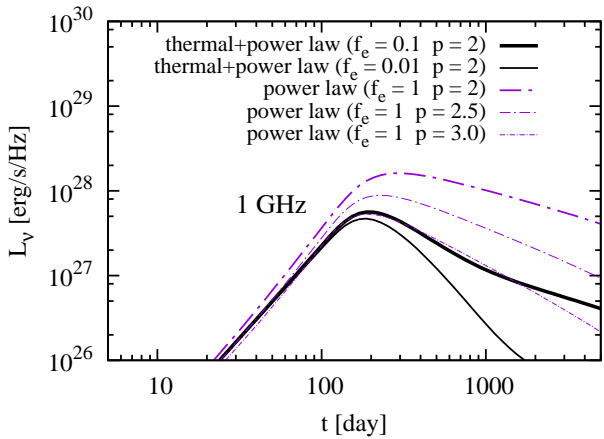}
\caption{ 
1 GHz light curve of a massive black hole formation with a mini-disk. 
The model parameters are the same as Fig. \ref{fig:fbt} except for the injection spectrum of electron at the forward shock. 
}\label{fig:injection_spec}
\end{figure}

Next we discuss the dependence of the radio emission on the phenomenological parameters of shock acceleration. 
In Fig.\ref{fig:B_amp}, we show radio light curves with different magnetic-field amplification efficiencies. 
With a larger $\varepsilon_B$, the synchrotron emissivity becomes larger and the peak flux becomes larger. 
A larger synchrotron emissivity at the same time leads to a larger SSA opacity, thus the peak times are delayed. 
In Fig.\ref{fig:injection_spec}, we compare 1 GHz light curves with different injection electron spectra.  
The black solid lines show the cases with thermal + power-law injection; a larger number fraction of electrons are accelerated in the thicker line, 
and dotted dash lines show the cases with pure power-law injection; a thicker line has a harder spectrum.
The peak time and flux so as the detectability of the radio transients are relatively insensitive to the injection spectrum. 
This is because that the SSA peak is predominantly determined by the typical electron Lorentz factor, which is $\sim \gamma_{\rm eT}$ for $f_{\rm e} \ll 1$ and $\sim \gamma_{\rm ei}$ for $f_{\rm e} \sim 1$. 
The detailed shape of the light curve depends on the injection spectrum; 
the rising part is proportional to $\propto t^{2}$ or $\propto t^{5/2}$ for a thermal or power-law dominated spectrum, respectively, 
while the decaying part is more shallower for a harder spectrum. 
In principle, it is possible to probe physics of trans-relativistic collisionless shock acceleration by identifying these features
although challenging given the anticipated signal-to-noise ratio and other uncertainties of radio observations.

\begin{figure}
\centering
\includegraphics[width=85mm]{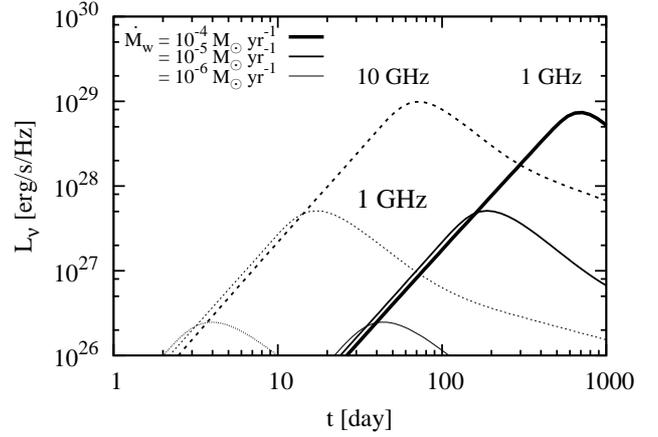}
\caption{ 
1 and 10 GHz light curve of a massive black hole formation with a mini-disk. 
The model parameters are the same as Fig. \ref{fig:fbt} except for the wind mass loss rate of progenitor star. 
}\label{fig:dotM}
\end{figure}

\begin{figure}
\centering
\includegraphics[width=85mm]{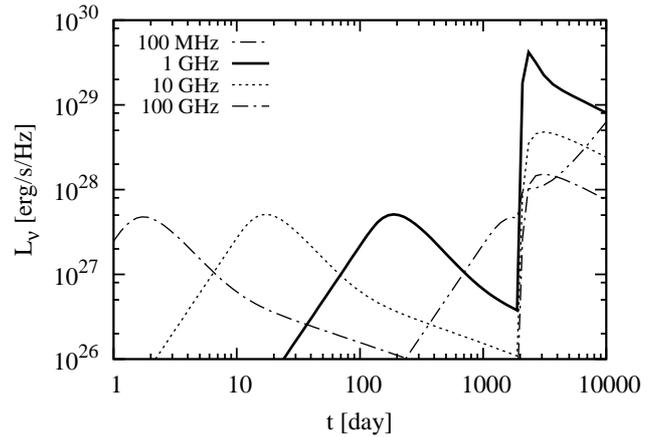}
\caption{ 
Same as Fig. \ref{fig:fbt} but with $n =10\ \rm cm^{-3}$ and a longer time evolution. 
A radio re-brightening can occur when the outflow reaches the outer edge of the wind bubble.
}\label{fig:ambient_structure}
\end{figure}

\begin{figure}
\centering
\includegraphics[width=85mm]{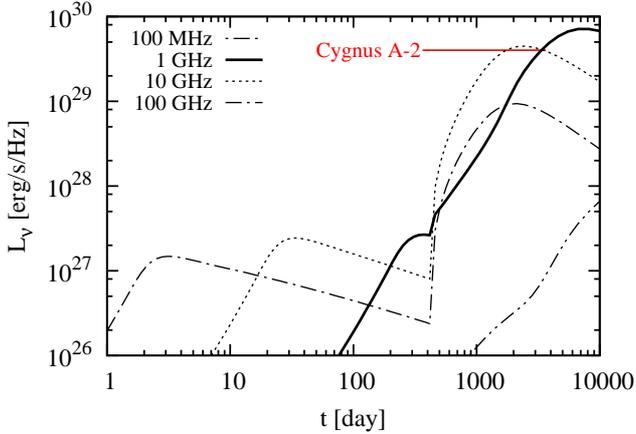} 
\caption{ 
Same as Fig. \ref{fig:ambient_structure} but with a slower ejecta velocity ($v = 0.15 \ c$), 
a denser cricumstellar medium ($n = 100 \ \rm cm^{-3}$, $\bar c_{\rm s}=3 \ \times 10^{7} \ \rm cm \ s^{-1}$) 
and an electron acceleration efficiency ($\varepsilon_{\rm B} = 0.33$, $f_{\rm e} = 0.1$).
The observed flux from Cygnus A-2 at $10 \ \rm GHz$ is also indicated.
}\label{fig:cygnusX2}
\end{figure}

The radio transients from the disk outflows are also sensitive to the ambient density structure. 
Fig. \ref{fig:dotM} shows 1 and 10 GHz radio light curves with different wind mass loss rates. 
A larger mass loss rate results in a denser wind medium. 
Then, a larger energy is dissipated at a fixed radius and the peak luminosity also becomes larger.
On the other hand, the SSA optical depth becomes large, thus the peak of the light curve at a given frequency is delayed. 
We note that the mass loss rate of massive stars is still uncertain especially in the pre-collapse phase and of great interest in the context of massive stellar evolution. 
In general, followup observations for a decade can probe the ambient density structure of the BH progenitor up to $\sim$ pc. 
Fig. \ref{fig:ambient_structure} shows a longer time evolution of a case with a relatively low wind mass loss rate. 
In such cases, the outflow is injected into the surrounding interstellar medium without significant deceleration, 
where the energy dissipation rate by the shock jumps up, resulting in forming the second peak in the radio light curves. 
These second peaks can be also detectable for events occurring at cosmological distances.

Fig. \ref{fig:cygnusX2} shows the result of another long-term calculation.
Here we choose the model parameters to explain the radio flux of Cygnus A-2~\citep{Perley_et_al_17}. 
Cygnus A-2, which was recently identified as a radio source at a projected offset of $460 \ \rm pc$ from Cygnus A, 
has a peak flux of $\sim (3-4) \times 10^{29} \ \rm erg \ s^{-1} \ Hz^{-1}$ at $\sim 10 \ \rm GHz$ and its variability timescale is $\sim 1-10 \ \rm yr$. 
These observed properties are broadly consistent with our model with a relatively high circumstellar density and electron acceleration efficiency.
The high denisty is feasible given that in Cygnus A-2 is likely in a highly star-forming near-nucleus region~\citep[e.g.,][]{Canalizo_et_al_03,Privon_et_al_12}.
Our model predicts that the flux at $\gtrsim 10 \ \rm GHz$ starts to decrease while the flux at $\lesssim \rm a \ few \ GHz$ increases in a few years, which can be tested by followup observations.

\section{Discussion}\label{sec:discussion}
We have studied black hole formation in collapsing massive stars, which is accompanied by a mini-disk with a mass of $\lesssim M_\odot$. 
Based on 1D stellar evolution calculations and some analytical estimates, we show that such a situation can occur in metal-poor massive stars and/or massive stars in close binaries. The accretion rate of the mini-disk is typically super-Eddington and an ultrafast outflow with a velocity of $\gtrsim 0.1\ c$ will be launched from the disk. 
Without significant quasi-spherical explosion in advance, the outflow is almost directly injected into the circumstellar medium. 
We have calculated synchrotron emission from relativistic electrons accelerated at the shock occurring between the outflow and circumstellar medium, and found that radio transients with a flux of $\sim 10^{26-30} \ \rm erg \ s^{-1} \ Hz^{-1}$ days to decades after the BH formation can be produced. 
The details of the light curve depends on the model parameters; the outflow velocity and mass, the electron acceleration efficiencies, and the circumstellar density profile, which can be probed by radio observations in principle.  
Although the uncertainties are large, the occurrence rate of the BH formation with a mini-disk may range from $\sim 0.1 \ \%$ to $\sim 10 \ \%$ of core-collapse SNe; the lower bound corresponds to the observed GRB and binary BH merger rates while the upper bound to the rate of failed SNe. 

Extragalactic radio transients with the above duration, flux, and event rate have been searched for. 
So far, the tightest constraints on the $\sim 1-10 \ \rm GHz$ radio transient rate have been obtained by the pilot observation for the Caltech-NRAO Stripe 82 Survey on timescales of a week to yrs~\citep{Mooley_et_al_16}, the FIRST Survey on timescales of minutes to a yr~\citep{Thyagarajan_et_al_11}, the FIRST and NVSS surveys targeting orphan GRB afterglows~\citep{Levinson_et_al_02}, and a 22-yr survey with the Molonglo Observatory Synthesis Telescope~\citep{Bannister_et_al_11}. 
Roughly, these surveys put constraints on the areal number density of radio transients as $N (> S) \lesssim 0.1 \ {\rm deg^2} \ (S/0.1 \ {\rm mJy})^{-3/2}$~\citep[see Fig. 22 of][]{Mooley_et_al_16}. 
Equivalently, the event rate of radio transients is inferred to be ${\cal R} (> S) \lesssim 0.1 \  {\rm deg^2 \ yr^{-1}} \ (S/0.1 \ {\rm mJy})^{-3/2} (\delta t/ 1 \ {\rm yr})^{-1}$, where $\delta t$ is the evolution timescale of the transients. 
In terms of the intrinsic luminosity per frequency, $L_\nu$, this can be rewritten as ${\cal R} (> L_\nu) \lesssim 10^{-4} \  {\rm Mpc^{-3} \ yr^{-1}} \ (L_\nu/10^{28} \ {\rm erg \ s^{-1} \ Hz^{-1}})^{-3/2} (\delta t/ 1 \ {\rm yr})^{-1}$.
Note that this value is comparable to the local core-collapse SN rate of $\sim 10^{-4} \  {\rm Mpc^{-3} \ yr^{-1}}$~\citep[see a review][and references therein]{Madau:2014bja}. In the coming years, surveys with e.g., the Jansky VLA~\citep{Perley_et_al_11}, ASKAP~\citep{Johnston_et_al_08}, MeerKAT~\citep{Booth_Jonas_12}, and Apertif/WSRT~\citep{Oosterloo_et_al_10} will provide a tighter constraint on the radio transient rate for a broader frequency range by orders-of-magnitude.

As shown in Fig. \ref{fig:cygnusX2}, interestingly, a proposed radio transient from newborn BHs with a mini-disk might already has been observed.  Cygnus A-2 resides in a highly star-forming region; the star formation rate of Cygnus A is as high as $\lesssim 100 \ M_\odot \ {\rm yr^{-1}}$~\citep{Privon_et_al_12} and the associated core-collapse SN rate can be $\lesssim 1 \ \rm yr^{-1}$. 
Cygnus A-2 has been observed with VLA and VLBA for $\sim 30 \ \rm yrs$, thus a lower limit on the event rate of Cygnus A-2 like transients can be a few $\%$ of core-collapse SNe, which is  consistent with our scenario. 

However, it may be difficult to distinguish the progenitors of radio transients only through radio observations. 
There are known extragalactic transient sources such as active galactic nuclei (AGNs), type I and II SNe, GRBs, tidal disruption events~\citep[see e.g.,][and references therein]{Mooley_et_al_16}, interaction-powered SNe~\citep{Murase_Thompson_Ofek_14}, and binary neutron star mergers~\citep[e.g.,][]{Nakar_Piran_2011,Hotokezaka_et_al_16,Hallinan+17}.
There are also theoretically proposed sources such as the accretion-induced collapse of white dwarfs~\citep[e.g.,][]{Piro_Kulkani_13,Moriya_16}, and binary BH mergers~\citep[e.g.,][]{Murase_et_al_16,Yamazaki_et_al_16}. 
In particular, AGNs and a brighter population of core-collapse SNe have a comparable flux and variability timescale but higher event rates compared with our scenario. 
Furthermore, even in a single source population, the emission model has a number of parameters; see Figs. \ref{fig:dotM} and \ref{fig:B_amp} for a parameter degeneracy between $\dot M_{\rm w}$ and $\varepsilon_B$. Multi-wavelength observations, including UV/optical measurements, are crucial to discriminate among the radio transients. 

In the case of BH formation with a mini-disk, a promising electromagnetic signal is the cooling emission from the outflow, which is observable as a luminous fast blue transient \citep[see Eqs. \ref{eq:t_p}-\ref{eq:Lbol_p} and][]{Kashiyama_Quataert_2015}. 
UV/optical transients with a duration of a few days is now being explored by e.g., PTF~\citep{Law_et_al_2009}, PanSTARRS~\citep{Hodapp_et_al_2004}, and Subaru HSC~\citep{Miyazaki_et_al_12}. Coordinated UV/optical and radio observations like the VLA all sky survey will be very much suitable for identifying newborn BHs with a mini-disk.  

It is worthwhile noting that similar UV/optical transients have been expected at the birth of massive BH binaries~\citep{Kimura+17}, which was also discussed in this paper. Close binaries involving a WR star may also lead to bright radio transients but with a larger kinetic energy caused by a more massive outflow. 

A fraction of the quasi-thermal photons can be inverse-Compton up-scattered by relativistic electrons at the forward shock. 
In fact, the IC emission is the dominant cooling process of the electrons at $t \lesssim t_{\rm peak}$ (Eq. \ref{eq:t_p}). 
Instead of Eqs. (\ref{eq:gamma_eM_syn}) and (\ref{eq:gamma_ec_syn}), the maximum energy Lorentz factor is given by $t_{\rm acc} = t_{\rm IC} = 3 m_{\rm e} c/(4 \sigma_{\rm T} a T_{\rm peak}^4 \beta \gamma_{\rm e})$, 
which yields $\gamma_{\rm eM} = [9eB\beta/(80\sigma_{\rm T} a T_{\rm peak}^4)]^{1/2}$, or 
\begin{equation}
\gamma_{\rm eM} \sim 6.4 \times 10^5 \ \left(\frac{T_{\rm peak}}{10^4 \ \rm K}\right)^{-2} \left(\frac{t_{\rm peak}}{1 \ \rm day}\right)^{-1/2}, 
\end{equation}
and the cooling Lorentz factor is estimated as $\gamma_{\rm ec} = 3m_{\rm e}c/(4 \sigma_{\rm T} a T_{\rm peak}^4 \beta t)$, or 
\begin{equation}
\gamma_{\rm ec} \sim 47 \left(\frac{T_{\rm peak}}{10^4 \ \rm K}\right)^{-4} \left(\frac{t_{\rm peak}}{1 \ \rm day}\right)^{-1},
\end{equation}
which suggests the slow-cooling regime. 
Assuming the Thomson regime, for $p\sim2$, the maximum IC luminosity is roughly estimated to be
\begin{eqnarray}
L_{\rm IC}^c&\sim&\frac{\epsilon_{\rm e}}{2\mathcal C}\frac{\dot{M}_w}{v_w}v_{\rm out}^3\nonumber\\
&\sim&6\times{10}^{40}~{\rm erg}~{\rm s}^{-1}~\epsilon_{\rm e,-1}{\mathcal C}_{1}^{-1}\dot{M}_{w,-5}v_{w,8}^{-1}v_{\rm out,10}^3,
\end{eqnarray}
where ${\cal C}$ is the bolometric correction factor and $\epsilon_e$ is the energy fraction carried by non-thermal electrons calculated from Eq. (\ref{eq:e_cons}). In the case of $p = 2$,  $1/{\cal C}= \ln[\gamma_{\rm eM}/\gamma_{\rm ei}]^{-1}$.
The energy of the IC photons is 
\begin{equation}
\varepsilon_{\rm IC}^{c} \approx 2 k_{\rm B} T_{\rm peak} \gamma_{\rm e}^2 \sim 1.5 \ {\rm keV} \ \left(\frac{\gamma_{\rm e}}{30}\right)^2 \left(\frac{T_{\rm peak}}{10^4 \ \rm K}\right). 
\end{equation}
at which the differential luminosity is estimated to be $\sim L_{\rm IC}^c{(\varepsilon_{\rm IC}/\varepsilon_{\rm IC}^c)}^{1/2}$ for $p=2$. 
The IC emission also fades away on a timescale of $\approx t_{\rm peak} \sim$ a few days, but can be still detectable by e.g., {\it Swift} XRT from $\lesssim \ 100 \ \rm Mpc$. 
Here we assume a sensitivity of $1\times10^{-13} \, \rm erg \,cm^{-2}\, s^{-1}$ in $10^4$ seconds for blind searches. 
If the UV/optical, X-ray, and radio counterparts are simultaneously detected, the magnetic field energy fraction at the forward shock ($\epsilon_B$) can be determined. 


\section*{Acknowledgements}
KK thanks Andrei Beloborodov for valuable discussions.
KK also thanks Eliot Quataert and Rodrigo Fernandez for stimulating discussions. 
This work is supported by KAKENHI 17K14248 (KK), Flatiron Fellowship at the Simons Foundation and Lyman Spitzer Jr. Fellowship (KH), and by Alfred P. Sloan Foundation and NSF Grant No. PHY-1620777 (KM).

\appendix

\section{Analytic formulae of radio synchrotron spectrum}
We can analytically calculate the synchrotron spectrum when the electron spectrum is a single power law.
In the slow-cooling case, the SSA optical depth is analytically expressed as $\tau_{\rm ssa}=\xi_p [e\rho r/(m_{\rm p} B \gamma_{\rm ei}^5)] \times (\nu/\nu_{\rm i})^{-(p+4)/2}$, where $\xi_p$ is the numerical coefficient depending on the index $p$ \citep[e.g.,][]{Murase_Thompson_Ofek_14}, and then the SSA frequency is estimated from $\tau_{\rm ssa} (\nu_{\rm ssa}) \approx 1$.
In the case of wind medium $(r < r_{\rm w})$, we have
\begin{eqnarray}\label{eq:nu_ssa_w}
\nu_{\rm ssa} &\sim& 13 \ {\rm GHz} \ {\mathcal C}_p f_{\rm e}^{2/(p+4)} \zeta_{\rm e}^{2(p-1)/(p+4)}  \varepsilon_{B, -2}^{(p+2)/2(p+4)} \notag \\ 
&\times& \dot M_{\rm w, -5}^{(p+6)/2(p+4)} v_{\rm w, 8}^{-(p+6)/2(p+4)} \beta_{-1}^{(4p-6)/(p+4)} t_6^{-1}, 
\end{eqnarray} 
where ${\mathcal C}_p$ is a numerical factor depending on $p$, e.g., ${\mathcal C}_p \approx 1.5, 1.0, 0.70$ for $p = 2.0, 2.5, 3.0$, respectively.  
On the other hand, in the case of uniform-density medium $(r > r_{\rm w})$, 
\begin{eqnarray}
\nu_{\rm ssa} &\sim& 0.042 \ {\rm GHz} \ {\mathcal C'}_p f_{\rm e}^{2/(p+4)}  \zeta_{\rm e}^{2(p-1)/(p+4)}  \varepsilon_{B, -2}^{(p+2)/2(p+4)} \notag \\ 
&\times& n_0^{(p+6)/2(p+4)} \beta_{-1}^{5p/(p+4)} t_8^{2/(p+4)},
\end{eqnarray} 
where ${\mathcal C'}_p = 1.4, 1.0, 0.73$ for $p = 2.0, 2.5, 3.0$, respectively.  
The slow-cooling synchrotron spectrum is described as
\begin{equation}
F_\nu \approx F_{\nu, \rm max}
\begin{cases}
 \left(\frac{\nu_{\rm ssa}}{\nu_{\rm i}}\right)^{-(p-1)/2} \left(\frac{\nu}{\nu_{\rm i}}\right)^{5/2}  & (\nu_{\rm i} < \nu < \nu_{\rm ssa}) \\
 \left(\frac{\nu}{\nu_{\rm i}}\right)^{-(p-1)/2}  & (\nu_{\rm ssa} < \nu < \nu_{\rm c}) \\
 \left(\frac{\nu_{\rm c}}{\nu_{\rm i}}\right)^{-(p-1)/2} \left(\frac{\nu}{\nu_{\rm i}}\right)^{-p}  & (\nu_{\rm c} < \nu < \nu_{\rm M}),
\end{cases}
\end{equation}
for the case of $\nu_{\rm i} < \nu_{\rm ssa} < \nu_{\rm c}$. The SSA peak flux is 
\begin{eqnarray}\label{eq:F_sa_w}
F_{\nu, \rm ssa} &\sim& 0.12 \ {\rm mJy} \ {\mathcal C'}_p^{-(p-1)/2} f_{\rm e}^{5/(p+4)} \zeta_{\rm e}^{5(p-1)/(p+4)} \notag \\ 
&\times& \varepsilon_{B, -2}^{(2p+3)/2(p+4)} \dot M_{\rm w, -5}^{(2p+13)/2(p+4)} v_{\rm w, 8}^{-(2p+13)/2(p+4)} \notag \\ 
&\times& \beta_{-1}^{(12p-7)/(p+4)} d_{\rm 27}^{-2},
\end{eqnarray}
for wind media and 
\begin{eqnarray}\label{eq:F_sa_uni}
F_{\nu, \rm ssa} &\sim& 9.5 \ {\rm \mu Jy} \ {\mathcal C'}_p^{-(p-1)/2} f_{\rm e}^{5/(p+4)} \zeta_{\rm e}^{5(p-1)/(p+4)} \notag \\ 
&\times& \varepsilon_{B, -2}^{(2p+3)/2(p+4)} n_0^{(2p+13)/2(p+4)} \notag \\ 
&\times& \beta_{-1}^{(14p+6)/(p+4)} t_8^{(2p+13)/(p+4)} d_{\rm 27}^{-2}.
\end{eqnarray}
for uniform-density media.

\bibliographystyle{mnras}
\bibliography{ref}

\end{document}